\documentclass[11pt]{article}

\usepackage[margin=1in]{geometry}
\usepackage[T1]{fontenc}
\usepackage[utf8]{inputenc}
\usepackage{times}
\usepackage[hyphens]{url}
\usepackage{graphicx}
\usepackage{natbib}
\usepackage{caption}
\usepackage{algorithm}
\usepackage{algorithmic}
\usepackage{booktabs}
\usepackage{amsmath}
\usepackage{amssymb}
\usepackage{amsfonts}
\usepackage{multirow}
\usepackage{microtype}
\usepackage[colorlinks=true,linkcolor=black,citecolor=black,urlcolor=blue]{hyperref}

\title{Modality Relevance is not Modality Utility:\\
Post-hoc Selective Modality Escalation for Cost-Aware Multimodal RAG}

\author{
Xue Li, Yiming Gai \\
Hangzhou International Innovation Institute, Beihang University\\
Hangzhou, China
}

\date{}

\begin{document}
\maketitle

\begin{abstract}
Multimodal retrieval-augmented generation (RAG) grounds a generator in evidence drawn from heterogeneous modalities---text, tables, and images. The dominant deployment choice is binary and made \emph{before} the model has tried to answer: either run a cheap text(+table) pipeline, or pay for an expensive vision--language model (VLM) over every image. Recent adaptive systems improve on this by selecting the modality or fidelity \emph{pre-retrieval}, from a question-conditioned predictor of which modality will be needed. We show that this is the wrong decision point. Through an oracle headroom analysis on MultiModalQA, we find that the relevance of a modality to a question is a weak predictor of whether that modality is actually \emph{needed to answer correctly}: a large fraction of questions whose gold support includes an image are nonetheless answerable from text and tables alone, and a pre-retrieval router that escalates on apparent visual relevance over-escalates substantially relative to an oracle. We propose \textbf{post-hoc selective modality escalation}: answer cheaply from text and tables, run a verifier on the (query, draft answer, evidence) tuple that localizes \emph{which modality} is missing, and pay for VLM evidence only there. A calibrated value-of-escalation router then decides whether the expected accuracy gain justifies the visual cost. On MultiModalQA, our router recovers the accuracy of an always-on VLM pipeline while issuing far fewer visual calls, and closes most of the gap to the oracle escalation rate. The result extends a routing-signal hierarchy established for retrieval depth and reasoning hops to a third axis---modality---under a single cost-aware selective-escalation view.
\end{abstract}

\section{Introduction}

Retrieval-augmented generation (RAG) grounds large language models in external evidence by retrieving content and conditioning generation on it. In the multimodal setting, that evidence is heterogeneous: a question may be answerable from a passage, from a table cell, from the content of an image, or from a combination. Production multimodal RAG systems face a basic cost asymmetry. Reading text and tables is cheap, but turning an image into something a language model can use requires a vision--language model (VLM), which is markedly more expensive in latency and compute. The default deployment pattern resolves this asymmetry rigidly: either restrict the pipeline to text(+table) and accept failures on visually grounded questions, or invoke a VLM over every candidate image and pay that cost on every query---including the majority that never needed it.

Recent adaptive multimodal RAG aims to spend the visual budget selectively. A prominent line decides \emph{which modality or fidelity to retrieve} from a question-conditioned success predictor, before any evidence is read or any answer is drafted~\citep{voila2026}. Agentic pipelines go further, planning modality-specific sub-queries and aggregating across retrieval branches~\citep{hmrag2025,r1router2025}, but at the cost of many model calls per question. The first family is attractive because it is cheap, but it commits to the visual decision at the earliest and least-informed point in the pipeline: \emph{before the system has observed any evidence or attempted an answer}.

This paper argues, and shows empirically, that the modality decision should be made \emph{after} a cheap attempt, not before. Our central observation is a modality analogue of a relevance--utility gap that has been documented for retrieval scores in single-hop RAG: \textbf{the relevance of a modality to a question is not the same as its utility for answering}. A question can be \emph{about} an entity that has an associated image and still be answerable entirely from the surrounding text and table evidence; conversely, a question with little surface visual cueing may genuinely require reading a chart. On MultiModalQA~\citep{talmor2021multimodalqa}, we quantify this with an oracle headroom analysis: a large fraction of questions whose annotated gold support includes an image are answered correctly by a text+table pipeline with no image access at all, and an oracle that escalates to the VLM only when it is actually needed escalates on far fewer questions than a relevance-driven pre-retrieval router would. Spending the visual budget by apparent relevance therefore wastes calls on already-saturated questions while still mislocating some that need help.

We propose \textbf{post-hoc selective modality escalation}. The system first answers from a cheap text+table context and produces a draft answer. A verifier then inspects the (query, draft, evidence) tuple and, rather than emitting a flat ``insufficient'' signal, \emph{attributes} the evidence gap to a modality---most importantly, whether the missing evidence is visual. Only for questions where the gap is localized to the image modality does the system pay for VLM evidence, which we inject as compact textual sidecars and use to regenerate the answer. Because even this localized escalation can be unnecessary, a calibrated value-of-escalation router makes the final decision by comparing the predicted accuracy gain of escalation against its visual cost, exposing a single tunable operating point along the accuracy--cost frontier.

This design is a deliberate continuation of a routing-signal hierarchy. In single-hop RAG, conditioning the retrieval-depth decision on a drafted answer and its evidence dominates conditioning on query-only or retrieval-time signals. In multi-hop RAG, localizing the missing evidence to a specific reasoning hop and escalating only there---and only when the question is not already saturated--- dominates whole-question escalation. We show the same structure holds for modality: \emph{post-hoc, localized, calibrated} escalation dominates \emph{pre-retrieval, whole-question, relevance-driven} escalation. Modality is the third axis of one cost-aware selective-escalation principle.

\paragraph{Contributions.}
\begin{itemize}
\item \textbf{A modality relevance--utility gap.} An oracle headroom analysis on MultiModalQA showing that modality relevance weakly predicts modality utility, and that a pre-retrieval relevance-driven router over-escalates relative to an oracle.
\item \textbf{Post-hoc selective modality escalation.} A method that answers cheaply, localizes the evidence gap to a modality with a verifier, and escalates to VLM evidence only where needed, governed by a calibrated value-of-escalation router that exposes an explicit accuracy--cost operating point.
\item \textbf{Empirical results under a strict protocol.} Under a matched-budget, multi-seed held-out comparison on MultiModalQA, post-hoc routing beats a strengthened calibrated pre-retrieval baseline and approaches a modality oracle from learned signals; a WebQA cross-check shows the accuracy benefit scales with the relevance--utility gap. We report honest call-count and latency costs throughout.
\end{itemize}

\section{Related Work}

\paragraph{Retrieval-augmented generation.} RAG grounds language models in retrieved evidence~\citep{lewis2020rag} and has matured into a family of adaptive methods that decide \emph{whether} and \emph{how much} to retrieve. Self-RAG~\citep{asai2024selfrag} emits reflection tokens to gate retrieval and critique generations; CRAG~\citep{yan2024crag} adds a retrieval-quality evaluator that triggers corrective search; Adaptive-RAG~\citep{jeong2024adaptiverag} routes by predicted question complexity. These operate within a single (text) modality and are orthogonal to \emph{which} modality to spend on; like ours, their gating signals are cheapest before the expensive step yet---paralleling our finding---most informative when conditioned on a draft. Long-context studies further motivate not over-stuffing context, since irrelevant evidence degrades reading~\citep{liu2024lostmiddle}.

\paragraph{Multimodal RAG.} A large body of recent work extends RAG to heterogeneous evidence, from early multimodal retrievers~\citep{chen2022murag} to strong vision-language backbones~\citep{radford2021clip,li2023blip2,qwen2vl2024}. Surveys document a rapid expansion of methods that index and fuse text, tables, images, and structured content~\citep{mmragsurvey2025,mragdesign2025}, and document-centric benchmarks stress retrieval and reasoning over visually rich material~\citep{unidocbench2025}. The dominant questions in this line are \emph{what} to index and retrieve and \emph{how} to fuse modalities. Our question is orthogonal and, in deployment, often more consequential: given a cost asymmetry between cheap text and expensive vision, \emph{when} is it worth paying for the visual modality at all. We treat fusion as fixed and study the spend decision.

\paragraph{Adaptive and cost-aware modality selection.} Closest to our work are methods that allocate the visual budget selectively. VOILA~\citep{voila2026} selects modality and fidelity \emph{before} retrieval from a question-conditioned success predictor, the strongest representative of pre-retrieval routing and our primary baseline. Agentic multimodal pipelines plan modality-specific sub-queries and aggregate across retrieval branches~\citep{hmrag2025,r1router2025}; they are powerful but issue many model calls per question. R1-Router in particular routes \emph{during} step-wise reasoning, deciding when and where to retrieve from the evolving reasoning state---so post-hoc, dynamic routing is not absent in this literature. Our contribution is narrower and complementary: a single, cheap, \emph{calibrated} escalation decision that conditions on a grounded draft answer and exposes an explicit accuracy--cost operating point, rather than an open-ended multi-call agent. We differ from pre-retrieval routers on the \emph{decision point} (answer cheaply, then decide), and from agentic routers on \emph{cost discipline} (one extra draft and one verifier call, not a reasoning loop). This is what lets us separate modality \emph{utility} from modality \emph{relevance} at deployment-relevant cost. A growing body of cost-aware RAG work shares our tradeoff framing: tiered multimodal fallback that escalates to richer context only when needed~\citep{multifinrag2025}, selective retrieval cast as routing over knowledge sources~\citep{selfroutingrag2025}, explicit accuracy-versus-access-cost objectives~\citep{costawareevidence2026}, threshold-based modular cost-aware retrieval~\citep{perfvscost2026}, and empirical retrieval-\emph{depth} routing~\citep{costawaredepth2026}. Our specific contribution within this space is the \emph{post-draft} decision point for the \emph{modality} axis, with a calibrated value model and a strict matched-budget evaluation.

\paragraph{Evidence sufficiency and verification.} A line of work characterizes whether retrieved context is \emph{sufficient} to answer and uses this signal to gate retrieval or abstention. Our verifier inherits this idea but applies it at the level of modality attribution, and---crucially---is used as a \emph{router}, never as an answer judge: it decides whether to pay for more evidence, not whether the answer is correct. Methodologically the design continues a selective-escalation hierarchy established for two other axes: conditioning a retrieval-\emph{depth} decision on a drafted answer dominates query-only routing, and localizing a missing reasoning \emph{hop} dominates whole-question escalation. The present work shows the same structure governs the \emph{modality} axis.

\section{The Modality Relevance--Utility Gap}
\label{sec:gap}

We first establish, empirically, that the relevance of a modality to a question is a weak predictor of whether that modality is needed to answer it. We use MultiModalQA~\citep{talmor2021multimodalqa}, whose questions carry gold modality annotations, and measure an \emph{oracle headroom}: how accurately a system answers when it is told, per question, the minimal modality set actually required.

Table~\ref{tab:headroom} reports the result on the full development set ($n{=}2{,}441$). A text+table pipeline with no image access answers $35.1\%$ correctly; always invoking the VLM over every image reaches $44.6\%$ but at the maximum visual cost ($100\%$ escalation, $2{,}233$ context tokens). The key row is the oracle: escalating to the VLM \emph{only} on questions whose gold support requires an image reaches $43.0\%$---within $1.6$ points of always-on vision---while escalating on just $39\%$ of questions and using less than half the context tokens ($1{,}020$). In other words, roughly $61\%$ of questions are \emph{modality saturated}: their visually annotated content is not needed, because the answer is recoverable from text and tables. (The oracle's average context is in fact \emph{lower} than text+table, because for text- or table-only questions it uses only the single required modality rather than always concatenating both.) An even stronger reference point is the \emph{utility} oracle that escalates only when vision strictly turns a wrong answer right: it needs escalation on just $12\%$ of questions and reaches $46.9\%$---above always-on vision, because added images sometimes distract the generator into errors. A router that spends the visual budget by apparent relevance therefore pays for vision on far more questions than necessary. This is the modality analogue of a relevance--utility gap and the headroom our method targets.

\begin{table}[t]
\centering
\small
\begin{tabular}{lccc}
\toprule
Policy & Acc & Esc.\ rate & Ctx.\ tok. \\
\midrule
Text+table only      & 0.351 & 0\%   & 1{,}174 \\
Always-on VLM        & 0.446 & 100\% & 2{,}233 \\
Oracle modality      & 0.430 & 39\%  & 1{,}020 \\
\bottomrule
\end{tabular}
\caption{Oracle headroom on MultiModalQA dev ($n{=}2{,}441$). The oracle reaches near always-on accuracy while escalating to vision on only $39\%$ of questions: most questions are modality saturated.}
\label{tab:headroom}
\end{table}

\begin{figure}[t]
\centering
\includegraphics[width=\columnwidth]{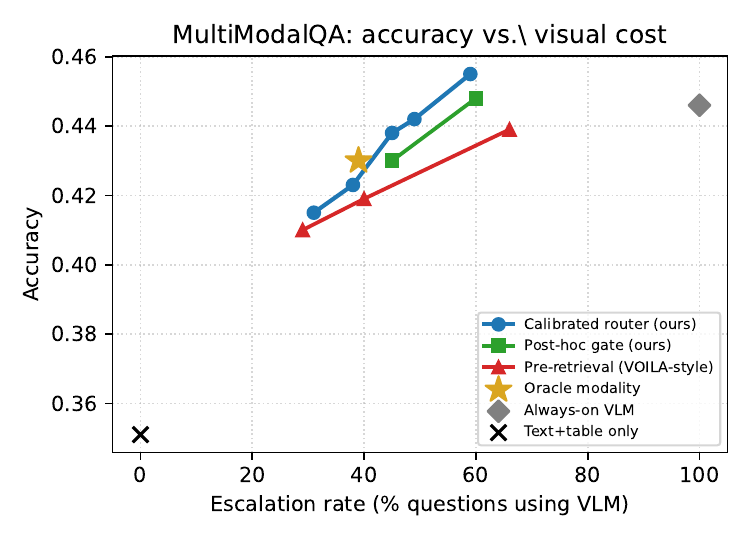}
\caption{Accuracy versus visual cost on MultiModalQA (operating points; see Table~\ref{tab:strict} for the strict held-out, multi-seed comparison). The calibrated router (ours) sits on or above the pre-retrieval frontier across the budget range and reaches close to the gold modality oracle ($\star$) from learned signals, approaching always-on vision ($\Diamond$) at far lower escalation.}
\label{fig:frontier}
\end{figure}

\section{Post-hoc Selective Modality Escalation}

\paragraph{Problem setup.} Given a question $q$ with a heterogeneous candidate pool of text, table, and image evidence, a modality policy chooses an action $a \in \{\textsc{keep}, \textsc{escalate}\}$. The \textsc{keep} action answers from a cheap text+table context $C_{\text{tt}}(q)$; the \textsc{escalate} action additionally pays to turn relevant images into language via a VLM and answers from the augmented context. Let $Y_a(q) \in \{0,1\}$ denote answer correctness under action $a$ and $\kappa(a)$ its context cost. A policy $\pi$ has accuracy $\bar A(\pi) = \mathbb{E}[Y_{\pi(q)}(q)]$ and escalation rate $\bar e(\pi) = \Pr[\pi(q)=\textsc{escalate}]$. We seek policies on the upper-left of the accuracy--escalation frontier.

\paragraph{Cheap draft and modality-gap verification.} The system first produces a draft answer $a_{\text{tt}}$ from $C_{\text{tt}}(q)$. A verifier then inspects the tuple $(q, a_{\text{tt}}, C_{\text{tt}})$ and, rather than emitting a flat sufficiency label, attributes any evidence gap to a modality---in our single visual modality, whether the missing evidence is an image. Only when the gap is attributed to the image modality does the system escalate: it generates compact textual sidecars for the top relevant images with a VLM and regenerates the answer. Crucially the verifier conditions on the draft answer, not merely on the question---this is what separates utility from relevance. A question whose images are topically relevant but whose answer is already correctly produced from text+table yields a draft that the verifier can accept, avoiding a needless visual call.

\paragraph{Calibrated value-of-escalation router.} A binary verifier commits to a single escalation rate. To expose the full frontier we instead learn calibrated estimates of per-action correctness from non-leaking post-draft features $\phi(q)$:
\begin{equation}
\hat p_{\textsc{keep}}(q) = \Pr[Y_{\textsc{keep}}=1 \mid \phi(q)], \quad
\hat p_{\textsc{esc}}(q) = \Pr[Y_{\textsc{esc}}=1 \mid \phi(q)],
\end{equation}
and escalate when the predicted value of escalation exceeds a threshold:
\begin{equation}
\pi_\tau(q) = \textsc{escalate} \iff \hat p_{\textsc{esc}}(q) - \hat
p_{\textsc{keep}}(q) \ge \tau .
\end{equation}
Sweeping $\tau$ traces the accuracy--escalation frontier with fine cost control, so the system can target any escalation budget---including rates below the binary verifier's fixed operating point. The features $\phi$ use only signals available after the cheap draft (draft abstention and length, question length, number of retrievable image candidates, the verifier's vote, and the keep-context cost) and exclude all gold annotations. The estimators are trained on counterfactual, selection-bias-free labels: for every training question we observe both the \textsc{keep} outcome and the \textsc{escalate} outcome.

\paragraph{Why a thresholded value rule is the right form.} The threshold rule is not an arbitrary heuristic but the optimal decision under a budget constraint. Consider maximizing expected accuracy subject to an escalation-rate budget $\bar e(\pi)\le B$:
\begin{equation}
\max_{\pi}\ \mathbb{E}\big[Y_{\pi(q)}(q)\big]\quad \text{s.t.}\quad \Pr[\pi(q)=\textsc{escalate}]\le B .
\end{equation}
Forming the Lagrangian with multiplier $\lambda\ge 0$, the per-question optimum escalates exactly when the expected accuracy gain exceeds $\lambda$, i.e.\ when $\mathbb{E}[Y_{\textsc{esc}}-Y_{\textsc{keep}}\mid \phi(q)]\ge\lambda$. Replacing the unknown conditional expectation with the calibrated estimate $\hat p_{\textsc{esc}}(q)-\hat p_{\textsc{keep}}(q)$ recovers the threshold rule above with $\tau=\lambda$. Thus the value-of-escalation difference, not the verifier vote, is the decision-relevant quantity, and $\tau$ is the price of one unit of budget; the verifier vote enters only as one feature of $\phi$. This is precisely why a router on the raw vote over-escalates: the vote tracks relevance, whereas optimality requires the conditional \emph{utility} difference.

\section{Experiments}

\paragraph{Setup.} We evaluate on MultiModalQA (full dev, $n{=}2{,}441$) and, for external validity, on WebQA~\citep{chang2022webqa}, using a balanced subset of $1{,}500$ questions ($750$ image-required, $750$ text-answerable). The answer generator is Qwen2.5-72B-Instruct; visual evidence is encoded once into compact textual sidecars by Qwen2.5-VL-7B and cached. We report string-match accuracy, escalation rate, and context tokens. Our main comparison uses a \emph{strict held-out, matched-budget} protocol: all routers share the same $50/50$ stratified split and the same calibration machinery; each is thresholded on the training split to hit a target escalation budget and evaluated once on the disjoint test split, averaged over $10$ seeds. This is the apples-to-apples setting a fair comparison requires.

\paragraph{Post-hoc vs.\ pre-retrieval, under a strict matched-budget protocol.} Our central result (Table~\ref{tab:strict}) compares three routers that differ \emph{only} in their feature set: random escalation, a \emph{calibrated} pre-retrieval router (question and image-candidate features only---a deliberately strengthened VOILA-style baseline given our exact calibration machinery), and our post-hoc router (which adds the draft answer and the modality-gap verifier vote). Both learned routers far exceed random escalation at every budget, confirming the gain is from the routing \emph{signal}, not the spend. Post-hoc routing beats the strengthened pre-retrieval baseline by a margin that \emph{grows with budget}: $+1.1$ points at $45\%$ ($0.442$ vs.\ $0.431$) and $+1.6$ at $50\%$, with per-seed standard deviation $\le 0.006$, so the advantage is small but stable rather than a single-split artifact. At the lowest budget ($30\%$) the two are within noise: when very few questions may escalate, the post-draft signal adds little. The honest reading is that post-draft conditioning helps most once there is enough budget to act on the questions it newly identifies.

\begin{table}[t]
\centering\small
\begin{tabular}{lccc}
\toprule
Target budget & Random & Pre-retrieval & Post-hoc (ours) \\
\midrule
30\% & $0.379$ & $0.424$ & $0.421$ \\
39\% & $0.386$ & $0.430$ & $0.435$ \\
45\% & $0.397$ & $0.431$ & $\mathbf{0.442}$ \\
50\% & $0.396$ & $0.431$ & $\mathbf{0.447}$ \\
\bottomrule
\end{tabular}
\caption{\textbf{Main result (held-out test, mean over $10$ seeds).} Strict matched-budget comparison; same split and calibration for all three, only the feature set differs. Per-seed std is $\le 0.011$ (full table in the appendix). Post-hoc routing dominates the strengthened pre-retrieval baseline, and the margin grows with budget.}
\label{tab:strict}
\end{table}

\paragraph{The verifier detects relevance, not utility.} Why does post-draft conditioning help, and where are its limits? Table~\ref{tab:verifier} evaluates the verifier's \texttt{need\_image} vote as a classifier against two labels: gold image-\emph{relevance} (image in the gold required modalities) and escalation \emph{utility} (escalating strictly turns a wrong answer right). The vote is an excellent relevance detector (F1 $0.87$, recall $0.94$) but a poor utility detector (F1 $0.32$). This is the relevance--utility gap made concrete: relevance is a predictable property of the question, but only $11.8\%$ of questions actually benefit from vision, so any router that escalates on the vote alone over-escalates. Our calibrated value model improves on the raw vote precisely because it predicts the keep-vs-escalate outcome \emph{difference} rather than relevance---and because the draft answer reveals whether text+table already sufficed.

\begin{table}[t]
\centering\small
\begin{tabular}{lcccc}
\toprule
Verifier vote vs.\ label & P & R & F1 & Acc \\
\midrule
Gold image-relevance & 0.80 & 0.94 & 0.87 & 0.89 \\
Escalation utility    & 0.21 & 0.78 & 0.32 & 0.62 \\
\bottomrule
\end{tabular}
\caption{The verifier vote is a strong \emph{relevance} detector but a weak \emph{utility} detector (MultiModalQA, full dev). Base rates: relevance $38.5\%$, utility $11.8\%$. This motivates a calibrated value model over the raw vote.}
\label{tab:verifier}
\end{table}

\paragraph{Oracle headroom and the calibrated frontier.} Table~\ref{tab:calib} reports the calibrated router across the full cost frontier (held-out test). Train-selected thresholds reproduce their target escalation rates on test to within about one point. The router reaches $0.423$ at $38\%$ escalation---close to the gold-support modality oracle ($0.430$ at $39\%$, a full-dev reference)---using only learned, non-leaking signals, and extends to $0.455$ at $59\%$. A stricter \emph{utility} oracle that escalates only when vision strictly helps needs just $12\%$ escalation and reaches $0.469$, \emph{above} always-on vision ($0.446$): images sometimes distract the generator into errors, so escalating everywhere is not optimal even ignoring cost.

\begin{table}[t]
\centering\small
\begin{tabular}{ccc}
\toprule
Target esc. & Test esc. & Test acc \\
\midrule
30\% & 31\% & 0.415 \\
39\% & 38\% & 0.423 \\
45\% & 45\% & 0.438 \\
50\% & 49\% & 0.442 \\
60\% & 59\% & 0.455 \\
\bottomrule
\end{tabular}
\caption{Calibrated router across the cost frontier (MultiModalQA, held-out test, $1{,}221$ questions). For reference (full dev): gold-support oracle $0.430$@$39\%$, utility oracle $0.469$@$12\%$, always-on vision $0.446$@$100\%$, text+table $0.351$@$0\%$.}
\label{tab:calib}
\end{table}

\paragraph{Cost and latency.} We report online query cost honestly (Table~\ref{tab:cost}), separating it from offline indexing. Sidecars are generated once and cached, so at query time no VLM call is made; the post-hoc method's overhead is instead the extra 72B draft and verifier calls, $2{+}\bar e$ per question ($2.45$ at $\bar e{=}0.45$) versus $1$ for the baselines, and an end-to-end latency of $4.49$\,s dominated by the verifier call. We therefore recommend the method where the goal is matched-budget accuracy or visual-call reduction, and note that delegating the verifier to a smaller model---an obvious mitigation we leave to future work---would close most of the latency gap.

\begin{table}[t]
\centering\small
\begin{tabular}{lccc}
\toprule
Method & 72B calls/q & Ctx.\ tok. & Latency (s) \\
\midrule
Text+table only & 1.00 & 1{,}174 & 0.91 \\
Always-on VLM   & 1.00 & 2{,}233 & 1.40 \\
Post-hoc (bal.) & 2.45 & 1{,}590 & 4.49 \\
\bottomrule
\end{tabular}
\caption{Honest online cost (MultiModalQA). Sidecars are precomputed offline; the post-hoc overhead is the extra draft and verifier calls, reported explicitly.}
\label{tab:cost}
\end{table}

\paragraph{External validity (WebQA).} As a cross-dataset sanity check, we run the full pipeline on WebQA, whose questions require \emph{exactly one} modality by construction, so relevance closely tracks utility and the gap we exploit is small. As expected, pre-retrieval and post-hoc routing converge in accuracy ($\sim$$0.355$ at $\sim$$49\%$ escalation) and the oracle ($0.351$@$50\%$) barely trails always-on vision ($0.356$); the calibrated router's only edge is cost control (it reaches sub-$49\%$ escalation, e.g.\ $0.356$@$44\%$). We read this not as a universal win but as evidence for the mechanism: \emph{the accuracy benefit of post-hoc escalation scales with the size of the relevance--utility gap}, and reduces to cost control when modalities are disjoint. A short ablation (appendix) shows the verifier vote is the most valuable router feature and that free-form sidecar descriptions carry the benefit.

\section{Limitations}
Our study has four main limitations. First, the answer model never consumes pixels: visual evidence reaches it as VLM-generated textual sidecars, so our claim is precisely that \emph{appending cached visual descriptions} helps selectively, not that the generator should attend to raw images; because sidecars are precomputed, the online VLM cost is an offline indexing cost, and a deployment that must caption on demand would shift the balance toward fewer escalations. Second, we consider a single expensive modality; extending modality-gap attribution to several competing expensive modalities (video, audio, tool-use tables) is future work, though the value-of-escalation formulation is agnostic to the number of actions. Third, the post-hoc design pays for a cheap draft and a verifier call on every question (Table~\ref{tab:cost}), so when vision is needed by nearly all questions the overhead is not amortized and an always-on policy is preferable---hence we state the accuracy claim \emph{conditionally}: post-hoc routing improves accuracy when cross-modal redundancy is substantial, and otherwise reduces to cost control. Fourth, the router is trained on offline counterfactual outcomes; deployment needs such logs or an exploration phase, and calibration may drift under distribution shift in question types or retrieval quality.

\section{Conclusion}
We argued that the modality decision in multimodal RAG should be made after a cheap attempt, not before it, because modality relevance is not modality utility---a gap we quantify directly: a verifier vote predicts gold relevance with F1 $0.87$ but escalation utility with F1 only $0.32$. A post-hoc verifier together with a calibrated value-of-escalation router beats a strengthened pre-retrieval baseline under a strict matched-budget held-out protocol, approaches a modality oracle from learned signals, and retains a cost-control advantage even when modalities are disjoint. We view modality as a natural third instance---alongside retrieval depth and reasoning hops---of one cost-aware principle: answer cheaply, verify against evidence, and escalate selectively only where it pays off; establishing that this is a single unifying law across all three axes, rather than a recurring empirical pattern, is left to future work.

\bibliographystyle{plainnat}
\bibliography{references}

\end{document}